\documentclass[sn-aps]{sn-jnl}% American Physical Society (APS) Reference Style
\usepackage{graphicx}%
\usepackage{multirow}%
\usepackage{amsmath,amssymb,amsfonts}%
\usepackage{amsthm}%
\usepackage{mathrsfs}%
\usepackage[title]{appendix}%
\usepackage{xcolor}%
\usepackage{textcomp}%
\usepackage{manyfoot}%
\usepackage{booktabs}%
\usepackage{algorithm}%
\usepackage{algorithmicx}%
\usepackage{algpseudocode}%
\usepackage{listings}%
\usepackage{bm}
\usepackage{braket}

\newcommand{\nn}{\nonumber}

\theoremstyle{thmstyleone}%
\theoremstyle{thmstyletwo}%

\theoremstyle{thmstylethree}%

\begin{document}

\title[Article Title]{Shape of $^{12}{\rm C}$ }

%%=============================================================%%
%% Prefix	-> \pfx{Dr}
%% GivenName	-> \fnm{Joergen W.}
%% Particle	-> \spfx{van der} -> surname prefix
%% FamilyName	-> \sur{Ploeg}
%% Suffix	-> \sfx{IV}
%% NatureName	-> \tanm{Poet Laureate} -> Title after name
%% Degrees	-> \dgr{MSc, PhD}
%% \author*[1,2]{\pfx{Dr} \fnm{Joergen W.} \spfx{van der} \sur{Ploeg} \sfx{IV} \tanm{Poet Laureate} 
%%                 \dgr{MSc, PhD}}\email{iauthor@gmail.com}
%%=============================================================%%

\author*[1]{\fnm{Masaaki} \sur{Kimura}}\email{masaaki.kimura@ribf.riken.jp}
\affil[1]{\orgdiv{Nishina Center}, \orgname{RIKEN}, \orgaddress{\city{Wako}, \postcode{351-0198}, \state{Saitama}, \country{Japan}}}
\author[2,1,3]{Yasutaka Taniguchi}
\affil[2]{\orgdiv{Department of Computer Science}, \orgname{Fukuyama University}, \orgaddress{\city{Fukuyama}, \postcode{729-0292}, \state{Hiroshima}, \country{Japan}}}
\affil[3]{\orgdiv{Department of Information Engineering}, \orgname{National Institute of Technology, Kagawa College (KOSEN)}, \orgaddress{\city{Mitoyo}, \postcode{769-1192}, \state{Kagawa}, \country{Japan}}}
%\affiliation{RIKEN Nishina Center, Wako, Saitama 351-0198, Japan}

\abstract{
We have examined the hypothesis by Bijker and Iachello who asserted that $^{12}{\rm C}$ has an internal structure with three $\alpha$ particles arranged in a triangular shape, leading to the formation of the ground rotational band consisting of $0^+$, $2^+$, $3^-$, $4^\pm$ and $5^-$ states. 
Following this idea, we reconstructed the intrinsic shape of $^{12}{\rm C}$ using experimental electron scattering data with minimal theoretical assumption. 
Our sole assumption was that the observed $0^+_1$, $2^+_1$, $3^-_1$, and $4^+_2$ states share a common internal structure, forming a rotational spectrum.
The reconstructed intrinsic density showed a beautiful triangular shape with three peaks implying  $\alpha$ cluster formation in the ground band. 
}


%%\pacs[JEL Classification]{D8, H51}
%%\pacs[MSC Classification]{35A01, 65L10, 65L12, 65L20, 65L70}
 
\maketitle 
 
\section{Introduction}\label{sec:intro}
$^{12}{\rm C}$ is one of the most studied nuclei in the last decades. The hypothesis that the Hoyle state (the $0^+_2$ state) is a Bose-Einstein condensate of three $\alpha$ particles~\cite{Tohsaki2001}, which was proposed by P.~Schuck and his collaborators, has ignited the interest in this nucleus and motivated so many studies. Today, not only the Hoyle state but also several excited states built on it are similarly understood as Bose-Einstein condensates~\cite{Funaki2009,Funaki2015,Schuck2016}, and experimental data for them has been accumulated~\cite{Itoh2011,Freer2012,Itoh2013a}. Furthermore, there have been significant advances in the exploration of 4$\alpha$~\cite{Wakasa2007,Funaki2008,Funaki2018,Adachi2018}, 5$\alpha$~\cite{Adachi2021,Zhou2023} and 5$\alpha$~\cite{Fujikawa2024} condensates and in the description of the Hoyle state by the first-principles calculations~\cite{Epelbaum2011,Epelbaum2012,Lovato2016,Otsuka2022,Shen2023}. 

\begin{figure}[tbp]
  \includegraphics[width=0.5\hsize]{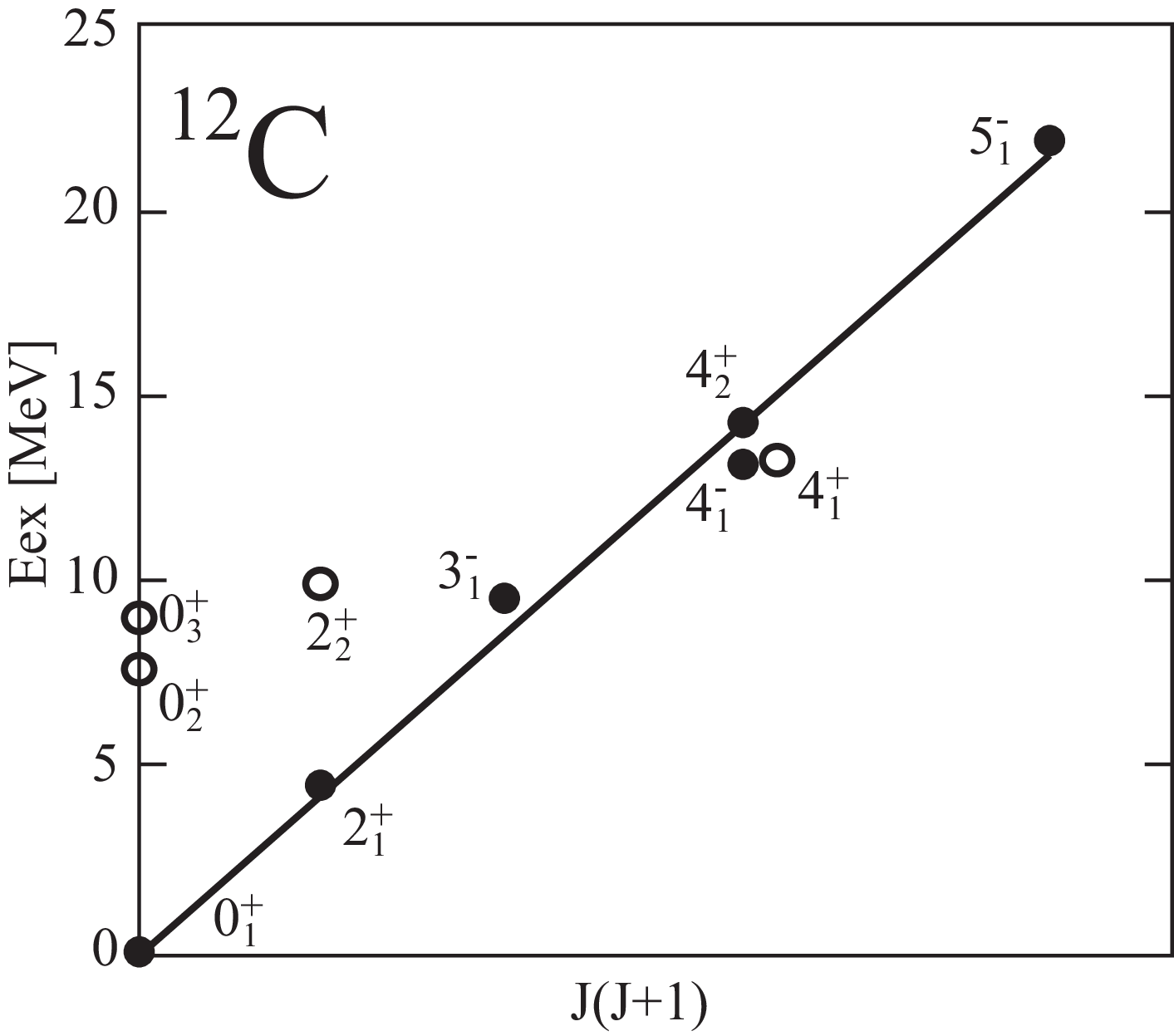}
   \caption{Partial level scheme of $^{12}{\rm C}$.  The proposed member states of the rotational ground band~\cite{Bijker2000, Marin-Lambarri2014} (the $0^+_1$, $2^+_1$, $3^-_1$, $4^+_2$, $4^-_1$, and $5^-_1$ states) are shown by filled circles, while the Hoyle state and the excited states associated with it~\cite{Funaki2015,Schuck2016} are shown by open circles.} 
   \label{fig:level}
\end{figure}

On the other hand, alternative explanations for the spectrum of $^{12}{\rm C}$ have been proposed from different perspectives~\cite{Stellin2016,Hess2018,Vitturi2020,Casal2021}. A notable example is the algebraic cluster model proposed by Bijker and Iachello~\cite{Bijker2000,Bijker2020}. This model asserts that the intrinsic state of $^{12}{\rm C}$ has a equilateral triangular arrangement of three $\alpha$ clusters. The ground band is attributed to the rotation of this intrinsic state, while other excited bands are explained by the vibration around the equilibrium shape coupled to the rotation. Although it cannot explain the observed transition density of the Hoyle state, the description of the ground band is intriguing. The rotation of the equilateral triangular shape yields a series of $0^+$, $2^+$, $3^-$, $4^\pm$, and $5^-$ states. An outstanding feature is the pair of the $4^+$ and $4^-$ states which fully degenerate due to the $D_{3h}$ symmetry. It is impressive to note that the observed $4^+_2$ (14.1 MeV) and $4^-_1$ (13.3 MeV) states are indeed almost degenerated (see Fig.~\ref{fig:level}). Moreover, the discovery of the predicted $5^-$ state (22.4 MeV)~\cite{Marin-Lambarri2014} revealed that  the $0^+_1$, $2^+_1$, $3^-_1$, $4^+_2$, $4^-_1$, and $5^-_1$ states manifest an approximate rotational spectrum, providing additional support for this hypothesis.

Combining these two hypotheses, one may be tempted to interpret the structure of $^{12}{\rm C}$  as follows: In the ground state, $\alpha$ particles are tightly bound and fixed at the vertices of a triangle. The rotation of this intrinsic structure gives rise to the ground band. When the excitation energy is used to break the bonding between $\alpha$ particles, they start to move freely, generating the gas-like states including the Hoyle state. It seems that the two hypotheses complementarily describe the structural transition of the $^{12}{\rm C}$ from a solid phase to a gas phase of $\alpha$ particles.

In order to support this interpretation, although the direct observation is not feasible, one may desire to reconstruct the internal structure of $^{12}{\rm C}$ in a model independent manner. Therefore, in this study, we visualize the shape of $^{12}{\rm C}$ from observables; the charge  and transition form factors measured by the electron scattering. We employ and extend an approach previously used for reconstructing the nuclear shape~\cite{Nakada1971,Kamimura1981}. As we want to discuss solely based on the experimental data, minimizing assumptions by theoretical models, the discussion here is limited to the ground band for which experimental data is available~\cite{Fregeau1956,Crannell1964,Crannell1966,Strehl1968,Sick1970,Nakada1971,Kline1973,Reuter1982,DeVries1987}. For the Hoyle state and its relatives, which are of priority interest, one needs to introduce a bit of theoretical calculations, and hence, we must revisit them in another paper.

\section{Numerical method}
For reconstructing the shape of $^{12}{\rm C}$, let us consider the electric transition density from the ground state to an excited state with the spin-parity of $(L^\Pi,L_z)=(L^\Pi,0)$. 
\begin{align}
  \rho^{0^+\rightarrow L^\Pi}(\bm r) &:= \braket{L^\Pi,0|\rho(\bm r)|0^+_1,0}
  = \braket{L^\Pi||\rho_L( r)||0^+_1}Y_{L0}(\hat r),\label{eq:transition_density_1}
\end{align}
where $\rho(\bm r)$ is the charge density operator, and $\hat{r}$ is the angles of the position vector $\bm r$ in polar coordinates. Note that the case of $L^\Pi=0^+_1$ corresponds to the charge density of the ground state. The last equality of this equation defines the reduced  transition density (reduced charge density for the $L^\Pi=0^+_1$ case), $\braket{L^\Pi||\rho_L(r)||0^+_1}$, through the Wiger-Eckart theorem. Their fourier transforms, the transition form factors $F_L(q)$, are observables. Within the plane wave impulse approximation, they are related to the elastic and inelastic electron-nucleus scattering cross sections as follows:
\begin{align}
 \frac{d\sigma^{\rm obs}_{L}}{d\theta} &= |F_L(q)|^2\frac{d\sigma_L^{\rm Mott}}{d\theta},
 \label{eq:form_factor_1}\\
 F_L(q) &:= \frac{\sqrt{4\pi(2L+1)}}{Z}\int_0^\infty r^2dr\ j_L(qr)\braket{L^\Pi||\rho_L(r)||0^+}, 
 \label{eq:form_factor_2}
\end{align}
where $j_L(qr)$ is the spherical Bessel function. $d\sigma^{\rm obs}_{L}/d\theta$ denotes the observed elastic and inelastic cross sections, whereas $d\sigma^{\rm Mott}_L/d\theta$  denotes the Mott cross section. Thanks to the experimental efforts~\cite{Fregeau1956,Crannell1964,Crannell1966,Strehl1968,Sick1970,Nakada1971,Kline1973,Reuter1982,DeVries1987}, accurate data for the transition form factors of the $2^+_1$, $3^-_1$, $4^+_2$ states ($L^\Pi=2^+$, $3^-$, $4^+$), as well as the charge form factor ($L=0^+$), are available.

\begin{figure}[htbp]
  \includegraphics[width=0.4\hsize]{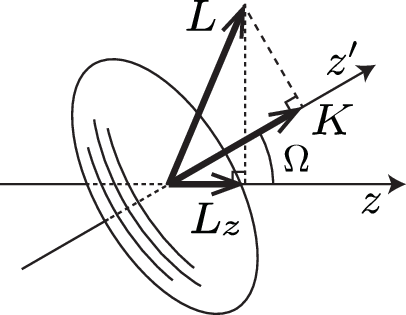}
   \caption{Schematic illustration of the rotational motion of the non-axial and parity-asymmetric rigid body.} 
   \label{fig:illust}
\end{figure}
In order to relate these observables to the shape of $^{12}{\rm C}$, we assume that the $0^+_1$, $2^+_1$, $3^-_1$ and $4^+_2$ states are the member of the ground band and share the same intrinsic state as asserted by Bijker and Iachello~\cite{Bijker2000,Bijker2020}. This implies that the wave functions of these states in the laboratory frame can be described as that of a non-axial and parity-asymmetric rigid rotor~\cite{Bohr1975} as illustrated in Fig.~\ref{fig:illust}.
\begin{align}
 \Psi_{L_zK}^{L^\Pi}&=\sqrt{\frac{2L+1}{16\pi^2(1+\delta_{K0})}}\chi(\xi,\Omega)
 \set{D^{L}_{L_zK}(\Omega) + \Pi(-)^{L+K}D^{L}_{L_z-K}(\Omega)},
 \label{eq:rigid_body_1}
\end{align}
where $\Omega$ and $D^L_{MK}(\Omega)$ are the Eular angles and Wigner's $D$ function,
which describe the collective rotation of rigid body. $K$ denotes the projection of $L$ onto the $z'$ axis of the intrinsic frame (see Fig.~\ref{fig:illust}). The internal structure of the rigid body, which is common to all the member states of the ground band, is described by $\chi(\xi,\Omega)$, where $\xi$ denotes the internal coordinates (the degrees-of-freedom of nucleons). Note that this is the only assumption we made in this study. 

Substituting this wave function into Eq. (\ref{eq:transition_density_1}) yields the equation that relates the transition densities to the density of the rigid body,
\begin{align}
  \rho^{0\rightarrow L^\Pi}(\bm r) &= \frac{1}{8\pi^2}
  \left(\frac{2L+1}{2(1+\delta_{K0})}\right)^{1/2} \nn\\
  &\times\int d\Omega\set{D^{L^*}_{L_zK}(\Omega) + \Pi(-)^{K}D^{L^*}_{L_z-K}(\Omega)}
  \braket{\chi(\xi,\Omega)|\rho(\bm r)|\chi(\xi,\Omega)}.
  \label{eq:transition_density_2}
\end{align}
To elucidate this relationship, let us define the density of the rigid body fixed at $\Omega=0$,
\begin{align}
  \rho^{\rm rigid}(\bm r) &:=  \braket{\chi(\xi,\Omega=0)|\rho(\bm r)|
 \chi( \xi,\Omega=0)}\nn\\
 &= \frac{1}{2}\sum_{lm}\rho^{\rm rigid}_{lm}(r)\set{Y_{lm}(\hat r)+(-)^mY_{l-m}(\hat r)}.
 \label{eq:rigid_density_1}
\end{align}
The second line of this equation defines the multipole decompositions of the rigid body density which satisfy the relation $\rho^{\rm rigid}_{lm}(r)=(-)^m\rho^{\rm rigid}_{l-m}(r)$. From Eq.~(\ref{eq:rigid_density_1}), the rigid body density at arbitrary angle $\Omega$ is given as,
\begin{align}
  \braket{\chi(\xi,\Omega)|\rho(\bm r)|\chi( \xi,\Omega)} &= R(\Omega)\rho^{\rm rigid}(\bm r)\nn\\
  &=\frac{1}{2}\sum_{lmm'}\rho^{\rm rigid}_{lm}(r)Y_{lm'}(\hat r)\set{D^l_{m'm}(\Omega)+(-)^m D^l_{m'-m}(\Omega)},
\end{align}
where $R(\Omega)$ is the rotation operator. Substituting this into Eq.~(\ref{eq:transition_density_2}) yields
\begin{align}
  \rho^{0\rightarrow L^\Pi}(\bm r) &= \frac{1}{\sqrt{2L+1}}\left(\frac{1+\Pi(-)^L}{1+\delta_{K0}}\right)^{1/2}\rho^{\rm rigid}_{LK}(r)Y_{L0}(\hat r),
\end{align}
and equating this with Eq.~(\ref{eq:transition_density_1}) produces the desired result.
\begin{align}
  \rho^{\rm rigid}_{LK}(r) &= \sqrt{2L+1}\left(\frac{1+\delta_{K0}}{1+\Pi(-)^L}\right)^{1/2}
  \braket{L^\Pi||\rho_L(r)||0^+_1}. \label{eq:rigid_density_2}
\end{align}
Thus, this equation relates the intrinsic density with the transition densities (observables), and  extends the previously used formula~\cite{Nakada1971,Kamimura1981} to the case of non-axial and parity-asymmetric shape.

To close this section, we summarize the procedure to reconstruct the shape of $^{12}{\rm C}$.
\begin{enumerate}
 \item The observed form factors are fitted by using Eq.~(\ref{eq:form_factor_2}) to extract the charge densities and electric transition densities $\braket{L^\Pi||\rho_L(r)||0^+_1}$ with $L=0^+_1$, $2^+_1$, $4^+_2$ and $3^-_1$, for which the experimental data is available.
 \item The multipole decomposition of the rigid body density, $\rho^{\rm rigid}_{LK}(r)$, are obtained from Eq.~(\ref{eq:rigid_density_2}).
 \item The rigid body density is reconstructed by Eq.~(\ref{eq:rigid_density_1}).
 \end{enumerate}
We remark again that the procedure is basically based on the experimental data. 
The only assumption is that the members of the rotational ground band, the $0^+_1$, $2^+_1$, $3^-_1$ and $4^+_2$ states, share the same intrinsic wave function.

\section{Numerical results and discussion}
\subsection{Fitting of the form factors}
\begin{figure}[hbt]
  \begin{center}
    \includegraphics[width=0.9\hsize]{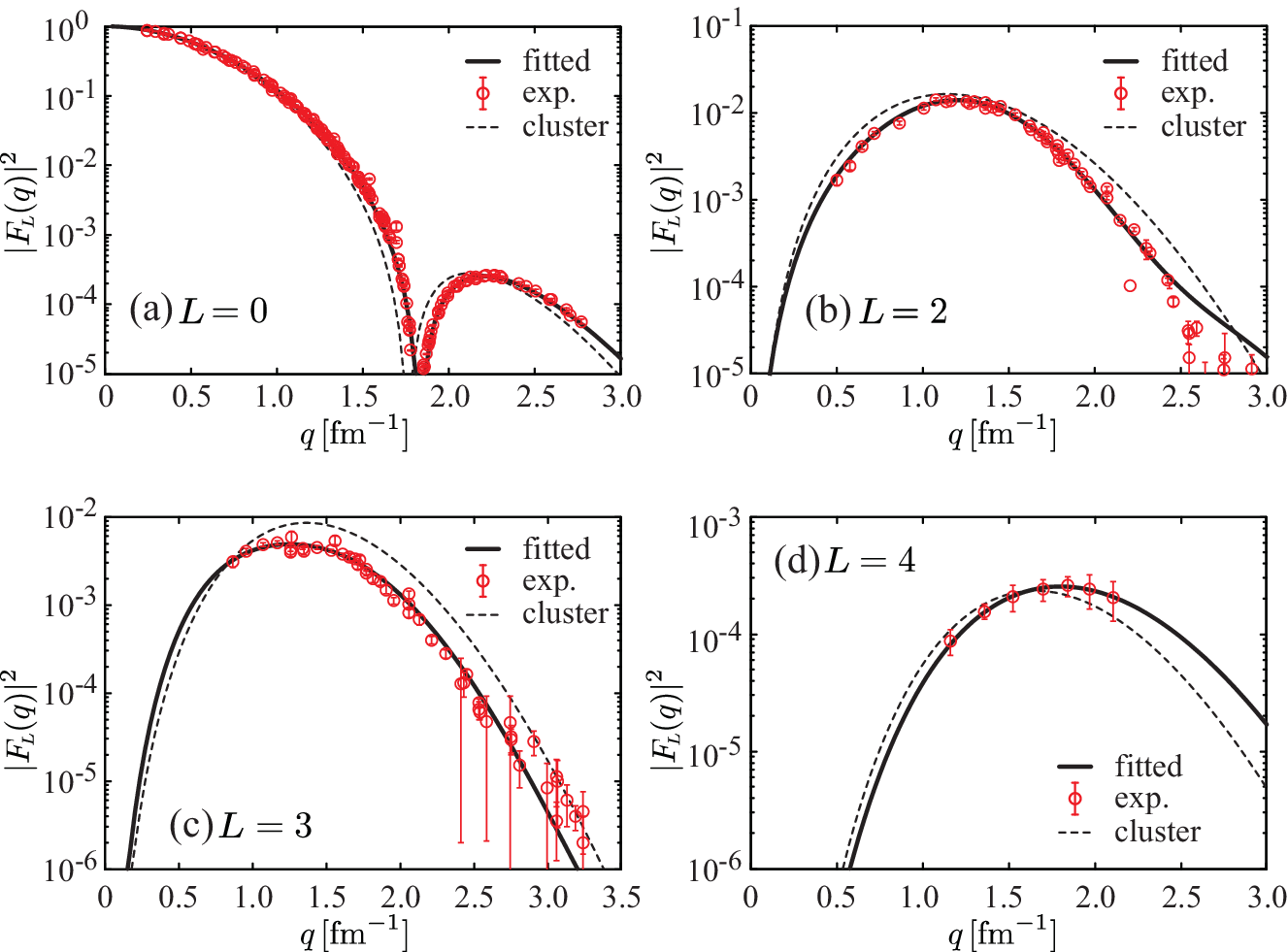}
    \caption{The observed (circles) and fitted (solid lines) charge form factor ($L=0^+$) and the electric transition form factors with $L=2^+$, $3^-$ and $4^+$. Those calculated by an $\alpha$ cluster model~\cite{Imai2019} (dashed lines) are also shown for comparison. } 
    \label{fig:form_factor}
  \end{center}
\end{figure}

Figure~\ref{fig:form_factor} compares the observed and fitted form factors with those calculated by an $\alpha$ cluster model~\cite{Imai2019}. 
In Eq.~(\ref{eq:form_factor_2}), we approximate $\braket{L^\Pi||\rho_L(r)||0^+_1}$ by the sum of the Gaussian functions as in Ref.~\cite{Kamimura1981},
\begin{align}
  \braket{L^\Pi||\rho_L(r)||0^+_1} \simeq \sum_{n=1}^{N}C_n\left(\frac{r}{R_n}\right)^L
  \exp\Set{-\left(\frac{r}{R_n}\right)^2}, \label{eq:sum_of_gaussian1}
\end{align}
where the parameters $C_n$ and $R_n$ are determined by the $\chi^2$-fit to the observed data. The number of the Gaussian is chosen as $N=10$ which is sufficient to reproduce the data.
The exceptional case is the $L=4$ form factor for which the number of the data points is not sufficient for the fitting. For this case, we first fit Eq.~(\ref{eq:sum_of_gaussian1}) to the form factor calculated by an $\alpha$ cluster model (dotted line in Fig.~\ref{fig:form_factor}). Then, we rescale the fitted result to minimize the $\chi^2$ to the experimental data points. As we see below, the contribution of the $L=4$ form factor to the reconstructed density is minor, and hence, the uncertainty in the fitting does not affect the results. 
We also tested the Fourier-Bessel expansion, which is another standard method to represent 
$\braket{L^\Pi||\rho_L(r)||0^+_1}$, and confirmed that the results are not affected by the choice of the representation.

\begin{figure}[tbp]
  \includegraphics[width=0.8\hsize]{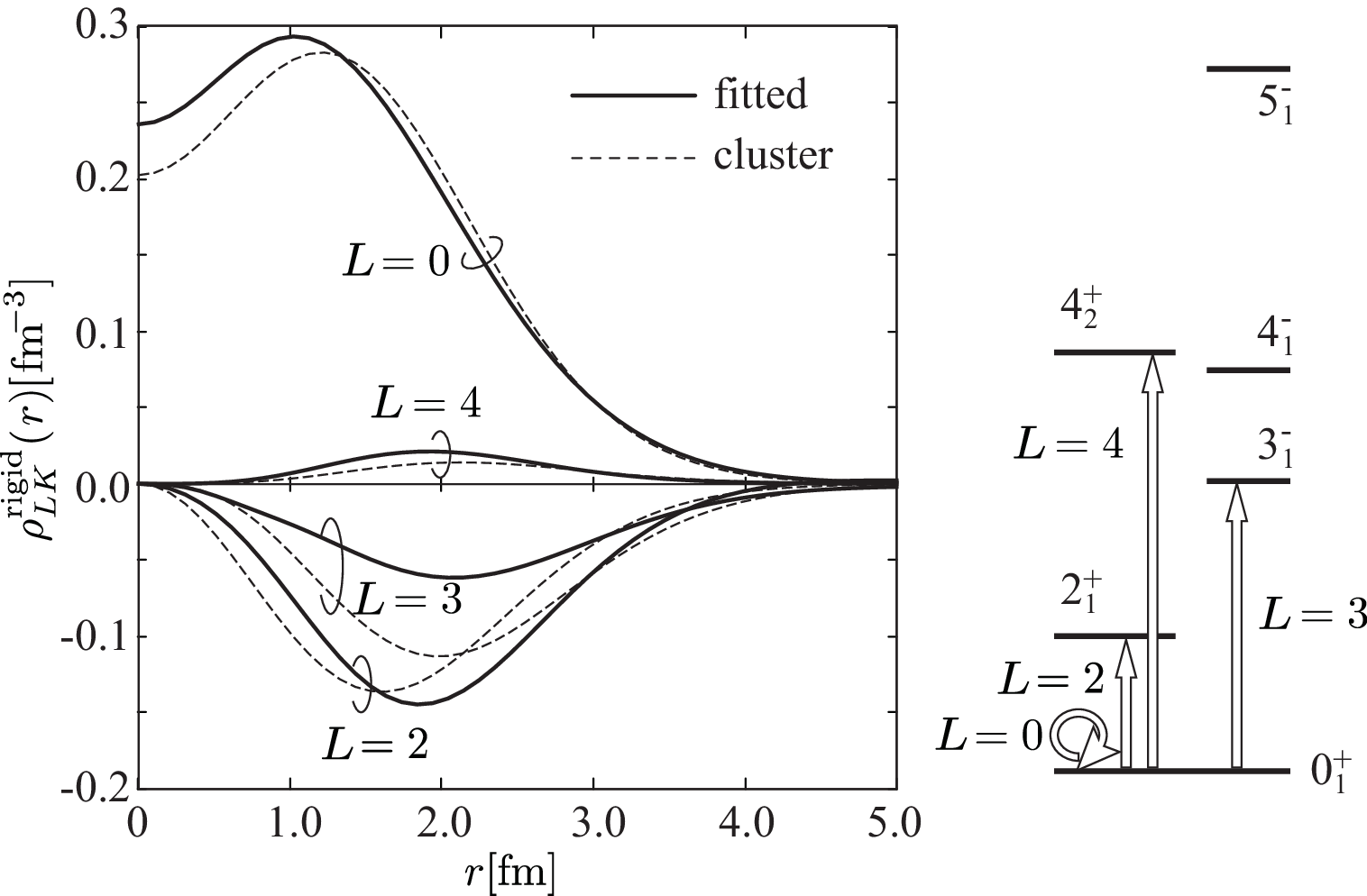}
   \caption{The multipole decomposition of the intrinsic density reconstructed from the observed form factors. The right panel shows the corresponding transition for each angular momentum $L$.} 
   \label{fig:density0}
\end{figure}

Figure~\ref{fig:density0} shows thus-obtained multipole decompositions of the intrinsic density in comparison with those calculated by an $\alpha$ cluster model. 
We have confirmed that the fitted densities reproduce the observed values of the charge radius~\cite{Reuter1982}, $B(E2)$~\cite{Pritychenko2016} and $B(E3)$~\cite{Crannell1967}. Hence, the fitted results look plausible\footnotemark[1].
\footnotetext[1]{Note that to convert the plots in Fig.~\ref{fig:density0} to densities, you need to multiply by spherical harmonics according to Eq.~(\ref{eq:transition_density_1}). For example, the central density of $L=0$ component is approximately $0.25/\sqrt{4\pi}\simeq 0.07\ \rm fm^{-3}$. Thus, combined with the neutron density, the central matter density is about $0.14\ \rm fm^{-3}$, slightly lower than the normal density.}

Here, we comment several uncertainties that arise in reconstructing the density. Firstly, as evident from Eq.~(\ref{eq:form_factor_1}), the sign of $F_L(q)$, hence that of $\rho^{\rm rigid}_{LK}(r)$, cannot be determined from the electron scattering experiments. 
However, the measured electric quadrupole moment of the $2^+_1$ state~\cite{Vermeer1983} allows us to uniquely determine the sign of the $L=2$ density to be negative. 
On the other hand, there is no way to determine the sign of the $L=4$ density from the existing data, but the choice of this sign has a limited impact on the reconstructed density since its amplitude is small. 
For the $L=3$ density, we will see that its sign has no influence on the results. 
We also comment that all the sign of the densities shown in Fig.~\ref{fig:density0} are, as far as we know, consistent with most of the theoretical calculations, which include the cluster models~\cite{Kamimura1981,Funaki2015,Imai2019}, Hartree-Fock~\cite{Ripka1968}, molecular dynamics models~\cite{Kanada-Enyo1998,Chernykh2007,Chernykh2010,Kanada-Enyo2007a}, and ab initio calculation~\cite{Lovato2016,Otsuka2022}.
The other uncertainty is in the choice of the quantum number $K$ of Eq.~(\ref{eq:rigid_density_2}). For the $L=2$ and $4$ densities, the choice of $K=0$ is natural since no $1^+$ and $3^+$ states exist below the $2^+_1$ and $4^+_2$ states. As for the $L=3$ density, the choice of $K=3$ is natural since the $3^-_1$ states is followed by a sequence of the $4^-_1$ and $5^-_1$ states, and no negative-parity state exists below the $3^-_1$ state.  

\subsection{Reconstruction of the intrinsic density}
\begin{figure*}[htbp]
  \includegraphics[width=\hsize]{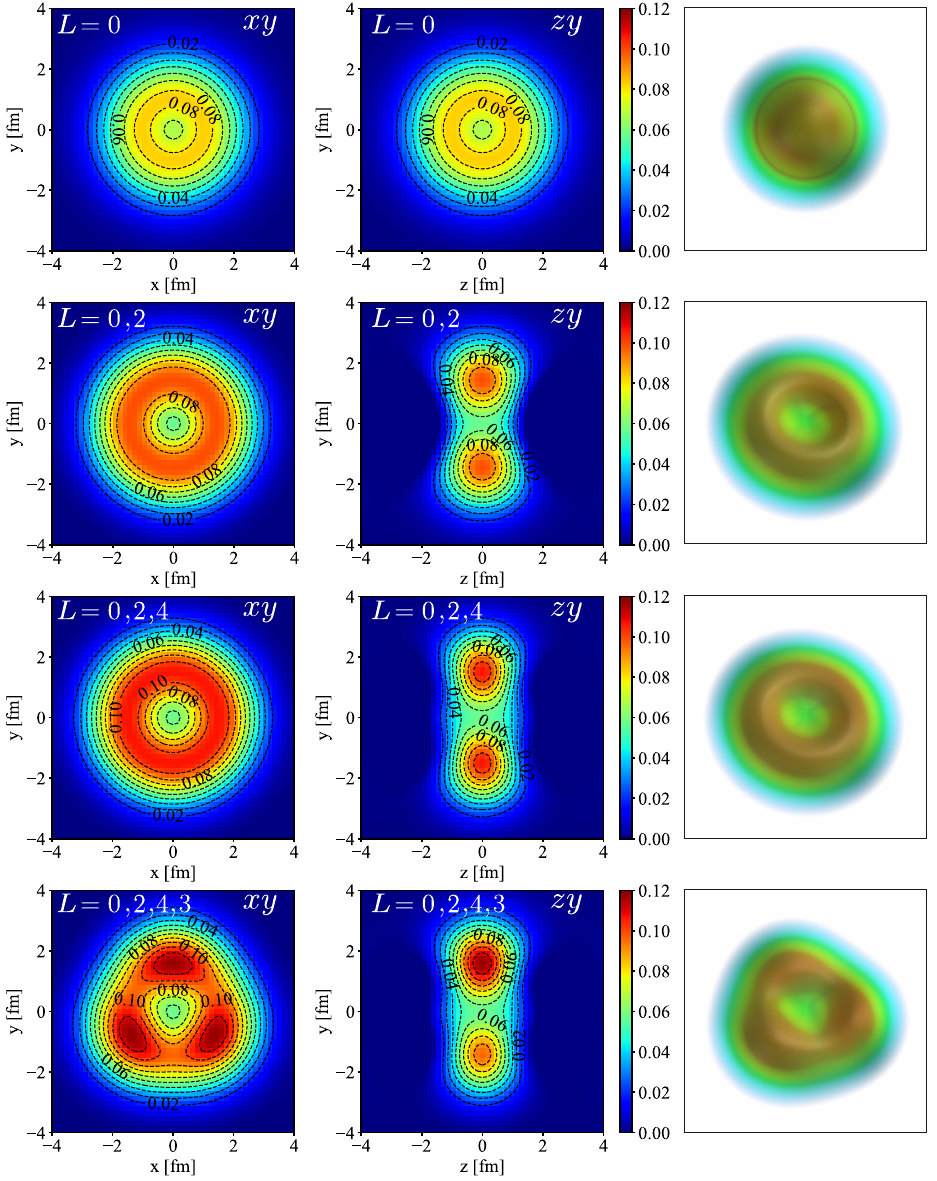}
   \caption{The intrinsic densities obtained from Eq.~(\ref{eq:rigid_density_1}). 
   Each row illustrates the reconstruction of density by summing multipole decompositions. The first row corresponds to the case where only the $L=0$ component is considered, and the last row represents the cumulative result where $L=0,2,4,3$ components are successively added.
   Each column, from left to right, illustrates the density distribution in the $xy$-plane, $zy$-plane, and three-dimensional space, respectively.}
   \label{fig:density1}
\end{figure*}

\begin{figure*}[tbp]
  \includegraphics[width=\hsize]{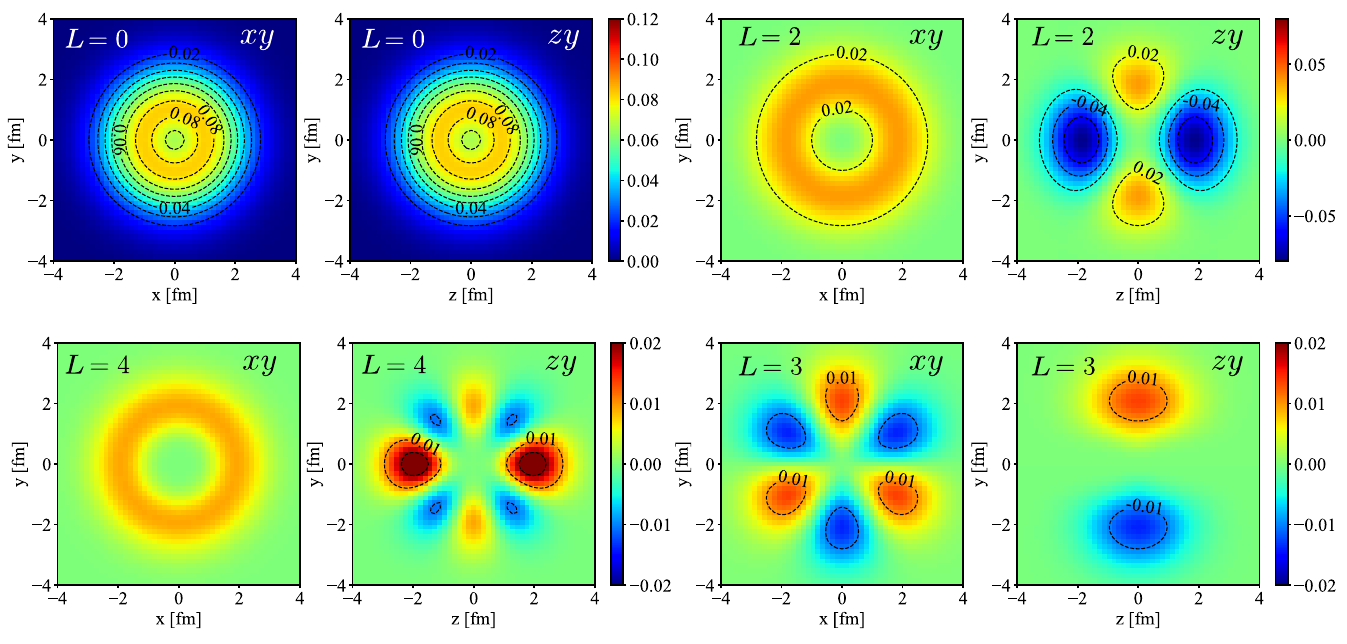}
   \caption{The decomposition of the intrinsic densitiy, ie. $\rho^{\rm rigid}_{LK}(r)\set{Y_{LK}(\hat r)+(-)^KY_{LK}(\hat r)}/2$, in the $xy$ and $zy$-planes. $K=0$ (K=3) for the $L=0,2,4$ ($L=3$) components. }
   \label{fig:density2}
\end{figure*}

According to Eq.~(\ref{eq:rigid_density_1}), Fig.~\ref{fig:density1} shows the reconstruction of the intrinsic density. The first row of the figure shows the result obtained by including only the $L=0$ component in Eq.~(\ref{eq:rigid_body_1}), naturally resulting in a spherical shape. A notable feature is a slightly reduced central density. A dip at $r=0$ originates in the change of sign of the form factor at 1.6 $\rm fm^{-1}$. From physical point of view, it is likely due to the larger number of protons occupying orbits other than the $s$-wave (approximately 4 protons) compared to those in the $s$-wave (approximately 2).

The second row shows the sum of the $L=0$ and $L=2$ components. Now the density distribution considerably deviates from spherical shape, showing a toroidal shape. This is attributed to the large negative value of the $L=2$ reduced transition density. As mentioned in the previous section, the magnitude and sign of the $L=2$ transition density are experimentally confirmed by the strong transitions probability, $B(E2)=4.65$ Wu.~\cite{Pritychenko2016}, and the electric quadrupole moment of the $2^+_1$ state, $Q=+6\pm3\ e\rm fm^2$~\cite{Vermeer1983}.  For reference, Fig.~\ref{fig:density2} illustrates each multipole decomposition of the intrinsic density from which one can imagine how a spherical shape transforms into a toroidal shape by the addition of the $L=2$ component.

The third row represents the sum of the $L=0$, 2 and 4 components. The toroidal shape is further emphasized by the $L=4$ component. As seen from Figs.~\ref{fig:density0} and \ref{fig:density2}, the amplitude of the $L=4$ component is not large, so its impact is limited.
Overlaying the $L=0$, 2, and 4 components is equivalent to regarding the $0^+_1$, $2^+_1$, and $4^+_2$ states as a series of rotational states, and this is not a novel idea. In fact, almost the same shape of $^{12}{\rm C}$ has already been obtained using experimental data~\cite{Nakada1971} and theoretical calculations~\cite{Kamimura1981}.

Finally, the last row of Fig.~\ref{fig:density2} shows the result of a novel idea which includes the $3^-_1$ state as a member of the rotational ground band.
Since we have chosen $K=3$ in Eq.~(\ref{eq:transition_density_2}), the $L=3$ component is proportional to $Y_{3\pm 3}(\hat r)$ and exhibits threefold symmetry about the $z$-axis, while $L=0$, 2, and 4 components are axially symmetric as shown in Fig.~\ref{fig:density2}. 
Consequently, the cumulative density also exhibits threefold symmetry implying the arrangement of three $\alpha$ particles at the vertices of an equilateral triangle. 

We remark that the maximum density of the spherical shape with $L=0$ was close to the normal density (see the left panel of Fig.~\ref{fig:density0} and the footnote in the previous section). As the $L\neq 0$ component reshaped it into a triangular form, the density at the vertex of the triangle becomes significantly high, approximately 0.11 $\rm fm^{-3}$. Including the contribution of neutrons, the matter density (protons plus neutrons) reaches around 0.22 $\rm fm^{-3}$, extremely higher than the normal density. Does this sound ridiculous? Among known nuclei, there is only one with such a high central density, and that is the $\alpha$ particle~\cite{Foris1987}.

\subsection{Discussions}
The reconstructed shape of $^{12}{\rm C}$ must have given rise to various questions. Here, we would like to mention some of them. 
Let us first consider why such an extremely deformed shape was obtained. Note that the threefold symmetry in the shape of $^{12}{\rm C}$ is not surprising. We have assumed $K=3$ when incorporating the $3^-_1$ state as a member of the ground rotational band. The threefold symmetry is merely a natural consequence of that assumption. The question here is why a transition from a spherical to a triangular shape occurred throughout the entire volume of $^{12}{\rm C}$.
For comparison, take a look at the shape of $^{152}{\rm Sm}$ (Fig.~7 in Ref.~\cite{Cardman1978}) reconstructed from electron scattering data in a same manner.
It is moderate and sharply contrasting with the present case. The extreme shape of $^{12}{\rm C}$ arises from the large amplitudes and unique spatial distribution of transition densities. This is evident when we compare the transition densities of $^{12}{\rm C}$  and $^{152}{\rm Sm}$ (see Fig.~4 of Ref~\cite{Cardman1978}, ). In $^{152}{\rm Sm}$, the transition densities with $L\neq 0$, which describe deviations from spherical shape, only have significant values near the nuclear surface, while being almost zero in the nuclear interior. As a result, deviations from spherical shape occur only at the nuclear surface, resulting in well-known prolate shape. In contrast, as seen in Fig.~\ref{fig:density0},  $^{12}{\rm C}$ has substantial amplitudes of the transition densities with $L\neq 0$ within the nuclear interior. Consequently, a deviation from spherical shape takes place across the entirety of the nucleus.

Next, let us consider a reasonable question of whether we can recognize the $3^-_1$ state (and subsequently the $4^-_1$ and $5^-_1$ states) as members of the ground rotational band.
We currently lack clear experimental evidence for this.
The abnormally large $B(E3)$~\cite{Crannell1967} value between the ground state and the $3^-$ state serves as indirect evidence indicating similar structure of these states. 
Little is known about the $4^-_1$ and $5^-_1$ states except for their energies. 
Therefore, the studies of $\alpha$-decay and electric transitions of these states will provide clues. In addition, investigating the transition form factor of the $5^-_1$ state is crucial, as it also contributes to the shape reconstruction.

Finally, we comment on the shape of the Hoyle state and its relatives. Reconstructing the shape of the Hoyle state requires the elastic and non-elastic scattering off the Hoyle state, but it is impractical. Moreover, the transition density between resonance states is not well defined. Note that the triangular model fails to reproduce the transition density of the Hoyle state, while the BEC model succeeds.
Therefore, the Hoyle state may not have definite shape.\footnotemark[2]
\footnotetext[2]{It is evident from the theoretical point of view~\cite{Tohsaki2001,Otsuka2022}, but it is not easy to prove it by experiments.}.

\section{Summary}
We reconstructed the density distribution of $^{12}{\rm C}$ from measured form factors by electron scattering, assuming that the $0^+_1$, $2^+_1$, $3^-_1$, and $4^+_2$ states share the same internal structure and form a rotational band. This attempt extends previous studies to the parity-asymmetric and non-axial symmetric cases. The reconstructed density exhibits a triangular shape, suggesting the formation of $\alpha$ clusters in the ground rotational band. This remarkable structure of $^{12}{\rm C}$ not only challenges traditional understanding of atomic nucleus but also poses several challenges to address, such as the question of whether the $3^-_1$, $4^-_1$, and $5^-_1$ states can be recognized as members of the ground rotational band.

\bmhead{Acknowledgments}
  The authors acknowledge the fruitful discussions with Dr. Q. Zhao, Dr. B. Zhou and Dr. Y. Funaki. They also thank to Dr. T. Neff and Dr. P. von Neumann-Cossel for the discussion and providing us the latest form factor data. 
  
  We dedicate this work to Prof. Peter Schuck, whose pioneering study of $\alpha$ particle condensation ignited our interest in nuclear cluster physics. One of the authors fondly remembers the discussion with him on the results presented here. It was just before the onset of COVID-19, marking our last conversation in person. Although he might not have fully endorsed this idea, it felt like a fitting topic to pay tribute to his lasting impact on our community.

\bibliography{sn-bibliography}% common bib file
%% if required, the content of .bbl file can be included here once bbl is generated
%%\input sn-article.bbl

\end{document}